\documentstyle[prb,twocolumn,aps]{revtex}
\newcommand{\s}{\sigma}
\newcommand{\bs}{\bar\sigma}
\begin{document}
\draft
\title{Equation of motion approach to the solution of Anderson model}
\author{Hong-Gang Luo,$^{1,2}$ \cite{luo} Zu-Jian Ying,$^1$ Shun-Jin Wang$^{1,2,3}$}
\address{1. Department of Modern Physics, Lanzhou University,
Lanzhou 730000, PR China\\ 2. Institut f\"ur Theoretische Physik,
Universit\"at Giessen, Giessen 35392, FR Germany\\ 3. Institute of Modern Physics, 
Southwest Jiaotong University, Chengdu 610031, PR China.}
\date{}
\maketitle
\begin{abstract}
Based on an equation of motion approach the single impurity Anderson model(SIAM) is 
reexamined. Using the cluster expansions the equations of motion of Green functions 
are transformed into the corresponding equations of motion of connected Green 
functions, which provides a natural and uniform truncation scheme. A factor of two 
missing in the Lacroix's approximation for the Kondo temperature is gained in the 
next higher order truncation beyond Lacroix's. A quantitative improvement in the 
density of states at the Fermi level is also obtained.
\end{abstract}
The single impurity Anderson model(SIAM)\cite{pw61} is one of the most fundamental 
and probably best understood models in the field of strongly correlated electronic 
systems. It was proposed to describe the properties of magnetic impurities in 
non-magnetic metallic hosts~(for a review see e.g. Ref.\ \onlinecite{ga74}). Yet, 
the SIAM has been used widely to imitate mixed valence and heavy fermion systems.
\cite{km84,g86,dn87,pj88} Since the SIAM was proposed, a variety of standard 
techniques have been applied to it and new methods have been developed to study 
its static and dynamical properties,  basically in the whole parameter space.~( 
For a recent review see, e.g. Ref.\ \onlinecite{ac93}.)

The method of equations of motion(EOM) of Green functions(GFs) is one of the most 
important tools to solve the model Hamiltonian problems in condensed matter physics.
One of its most appealing features is that it can work in the whole parameter space.
Many authors have applied the EOM approach to the SIAM for finite 
$U$\cite{pw61,al91,km95} or infinite $U$.\cite{ap69,cy76,la81,la82,jp93} 
In order to close the chain of equations it is usual to introduce the conventional 
Tyablikov decoupling scheme.\cite{dnz60} Lacroix employed thus decoupling scheme 
for higher order GFs. \cite{la81,la82} In the limit of strong intra-atomic Coulomb 
interaction he did not get the correct expression of Kondo temperature
from his high- and/or intermediate-temperature solutions\cite{la81} at 
his approximation level.

In the present paper we shall go  beyond Lacroix's approximation in a slightly 
different way. In the following we use the correlation dynamics 
approach\cite{wang85} to treat the problem. After writting down the hierarchy of 
EOM of the GFs the Tyablikov decoupling scheme is not applied directly. Instead 
systematic cluster expansions are employed, which express the higher order GFs in 
terms of same order connected GFs and lower order GFs, the hierarchy of EOM of the 
usual GFs is thus transformed into that of EOM of the connected GFs. The connected 
GFs are defined such that they can not be reduced to the low order ones by any way 
of decoupling. The hierarchy of equations of the connected GFs provides a natural 
and uniform truncation scheme. Our formalism, which is essentially equivalent to 
the Tyablikov decoupling scheme, is more systematic. The well known results like 
mean field theory,\cite{pw61} Hubbard-I approximation,\cite{cy76} two-peak 
solutions,\cite{ac66} and Lacroix's results\cite{la81,la82} are recovered exactly 
in successively higher levels of truncation. To go beyond Lacroix we introduce 
even higher order truncation. The Kondo temperature obtained is in agreement with 
the exact one\cite{fh78} except for a prefactor. The density of states(DOS) of the 
SIAM is calculated numerically at finite temperature. A quatitative improvement on 
the DOS at the Fermi level is also obtained.

Consider the Hamiltonian of the conventional SIAM\cite{pw61}
\begin{eqnarray}
H &=&\sum_{k\s}\varepsilon_{k}\hat{n}_{k\s} + \sum_{\s}\varepsilon_{d}
\hat{n}_{d\s}\nonumber\\ &&+ \frac12U\sum_{\s}\hat{n}_{d\s}\hat{n}_{d\bs} + 
V\sum_{k\s}(c_{k\s}^{+}d_{\s}+d_{\s}^{+} c_{k\s}). \label{eq1}
\end{eqnarray}
The notation used is as usual. In this paper the DOS for conduction electrons is 
taken to be a constant, $\rho(\varepsilon)=\frac{1}{2D}$ as $-D<\varepsilon<D$. 
The width of the virtual bound state is $\Delta=\pi\langle V^2\rangle/(2D)$.
\cite{pw61}

Using the Zubarev notation \cite{dnz60} $G_{A,B}(\omega) = \ll A; B\gg$, it is 
straightforward to write down the EOM for the one-particle GFs
\begin{eqnarray}
\left(\omega-\varepsilon_{d}- \sum_k\frac{V^2}{\omega - \varepsilon_k}\right)&& \ll 
d_{\s};d_{\s}^{+}\gg =1\nonumber\\ && + U\ll\hat{n}_{d\bs}d_{\s};d_{\s}^{+}\gg, 
\label{eq2}
\end{eqnarray}
and for the high order GFs
\begin{mathletters}
\label{eq3}
\begin{eqnarray}
&&(\omega+\varepsilon_{d}-\varepsilon_{k'}- \varepsilon_{k}) \ll 
d_{\bs}^{+}c_{k'\bs}c_{k\s}; d_{\s}^{+}\gg = \nonumber\\&& -U\ll 
\hat{n}_{d\s}d_{\bs}^{+}c_{k'\bs}c_{k\s}; d_{\s}^{+}\gg + V(\ll 
d_{\bs}^{+}c_{k'\bs}d_{\s}; d_{\s}^{+}\gg\nonumber\\ && +\ll \hat{n}_{d\bs}c_{k\s}; 
d_{\s}^{+}\gg - \sum_{k''}\ll c_{k''\bs}^{+}c_{k'\bs}c_{k\s};d_{\s}\gg),
\label{eq3a}\\ &&(\omega+\varepsilon_{k'}-\varepsilon_{d}- \varepsilon_{k})\ll 
c_{k'\bs}^{+}d_{\bs}c_{k\s}; d_{\s}^{+}\gg = \nonumber\\&& U\ll 
\hat{n}_{d\s}c_{k'\bs}^{+}d_{\bs}c_{k\s}; d_{\s}^{+}\gg + V(\ll 
c_{k'\bs}^{+}d_{\bs}d_{\s}; d_{\s}^{+}\gg\nonumber\\ &&-\ll \hat{n}_{d\bs}c_{k\s}; 
d_{\s}^{+}\gg + \sum_{k''}\ll c_{k'\bs}^{+}c_{k''\bs}c_{k\s};d_{\s}^{+}\gg), 
\label{eq3b}\\ &&(\omega+\varepsilon_{k'}-\varepsilon_{k} - \varepsilon_{d})\ll 
c_{k'\bs}^{+}c_{k\bs}d_{\s}; d_{\s}^{+}\gg = \langle c_{k'\bs}^{+}c_{k\bs}\rangle 
\nonumber\\ &&+ U\ll \hat{n}_{d\bs}c_{k'\bs}^{+}c_{k\bs}d_{\s}; d_{\s}^{+}\gg +
V(\ll c_{k'\bs}^{+}d_{\bs}d_{\s}; d_{\s}^{+}\gg \nonumber\\ &&  -\ll 
d_{\bs}^{+}c_{k\bs}d_{\s};d_{\s}^{+}\gg + \sum_{k''}\ll 
c_{k'\bs}^{+}c_{k\bs}c_{k''\s}; d_{\s}^{+}\gg). \label{eq3c} \end{eqnarray}
\end{mathletters}
To save space, the EOM for the high order GFs $\ll \hat{n}_{d\bs}d_{\s};
d_{\s}^{+}\gg,~\ll \hat{n}_{d\bs}c_{k\s};d_{\s}^{+}\gg,~\ll 
d_{\bs}^{+}c_{k\bs}d_{\s};d_{\s}^{+}\gg,~\text{and}~ \ll c_{k\bs}^{+}d_{\bs}d_{\s}; 
d_{\s}^{+}\gg$ are omitted. One can find their EOM in Ref.\ \onlinecite{la81}. 
Instead of employing directly the Tyablikov decoupling scheme, we make use of a 
cluster expansions to separate the
connected part of the GFs. As an example, the high order GF $\ll 
\hat n_{d\bs}d_{\s}; d_{\s}^+\gg$ is expressed as follows:
\begin{eqnarray} 
\ll \hat n_{d\bs}d_\s,d_\s^+\gg = n_{d\bs}\ll d_\s,&&d_\s^+\gg 
\nonumber\\ &&+ \ll \hat n_{d\bs}d_\s,d_\s^+\gg_c, \label{eq4} 
\end{eqnarray} 
where $\ll...\gg_{c}$ represents a connected GF and $n_{d\bs} = \langle 
\hat n_{d\bs}\rangle$.  It is straightforward to derive the EOM for the 
connected GFs. We write down the first two as follows:
\begin{mathletters}
\label{eq5}
\begin{eqnarray}
&&\left(\omega-\varepsilon_{d} - Un_{d\bs}-\sum_{k}\frac{V^2}{\omega-\varepsilon_k} 
\right)\ll d_{\s};d_{\s}^{+}\gg = 1 
\nonumber\\&&~~~~~~~~~~~~~~~~~~~~~~~~~~~~~~~~~~~~+ U\ll \hat n_{d\bs}d_{\s};
d_{\s}^{+}\gg_{c},\label{eq5a}\\ &&[\omega-\varepsilon_{d}-U(1-n_{d\bs})]\ll 
\hat n_{d\bs}d_{\s}; d_{\s}^{+}\gg_{c} \nonumber\\&&= Un_{d\bs}(1-n_{d\bs})\ll 
d_{\s};d_{\s}^{+}\gg +V\sum_{k}(\ll \hat n_{d\bs}c_{k\s};d_{\s}^{+}\gg_{c}
\nonumber\\&&~~~~~~~+\ll d_{\bs}^{+}c_{k\bs}d_{\s}; d_{\s}^{+}\gg_{c} -\ll 
c_{k\bs}^{+}d_{\bs}d_{\s}; d_{\s}^{+}\gg_{c}).\label{eq5b} \end{eqnarray} 
\end{mathletters}

It is not difficult to obtain the EOM of the other connected GFs such as $\ll 
\hat n_{d\s}c_{k\s};d_{\s}^+\gg_c, \ll c_{k\bs}^{+}d_{\bs}d_{\s};d_{\s}^{+}\gg_c, 
\ll d_{\bs}^{+}c_{k\bs}d_{\s}; d_{\s}^{+}\gg_{c}, \ll d_{\bs}^{+} c_{k'\bs} c_{k\s};
d_{\s}^{+}\gg_{c},  \ll c_{k'\bs}^{+}d_{\bs} c_{k\s};d_{\s}^{+}\gg_{c}, 
~\text{and}\\ \ll c_{k'\bs}^{+}c_{k\bs}d_{\s};d_{\s}^{+}\gg_{c}$.

In the following we discuss briefly the solutions at different orders of 
approximation. The lowest order truncation, i.e.  $\ll \hat n_{d\bs}d_{\s};
d_{\s}^{+}\gg_{c} \approx 0$, leads to the mean field theory, while next 
higher order truncation, namely, \\ $\ll c_{k\bs}^{+}d_{\bs}d_{\s}; 
d_{\s}^{+}\gg_{c} \approx 0, \ll d_{\bs}^{+}c_{k\bs}d_{\s};d_{\s}^{+}\gg_{c} 
\approx 0, ~\text{and}~ \\ \ll \hat n_{d\bs}c_{k\s};d_{\s}^{+}\gg_{c} \approx 0$, 
corresponds to Hubbard-I approximation. Subsequently, the solution with two peaks 
localized at $\varepsilon_{d}$ and $\varepsilon_{d}+U$ with weights $1-n_{d\bs}$ 
and $n_{d\bs}$ respectively, can be obtained from the truncation $\ll 
c_{k\bs}^{+}d_{\bs}d_{\s};d_{\s}^{+}\gg_{c} \approx 0, \ll 
d_{\bs}^{+}c_{k\bs}d_{\s};d_{\s}^{+}\gg_{c}\approx 0, \ll 
d_{\bs}^{+}c_{k'\bs}c_{k\s};d_{\s}^{+}\gg_{c}\approx 0,~ \text{and}~ 
\ll c_{k'\bs}^{+}d_{\bs}c_{k\s}; d_{\s}^{+}\gg_{c}\approx 0$. To reach  
Lacroix's approximation,\cite{la81} the connected GFs like $\ll 
d_{\bs}^{+}c_{k'\bs}c_{k\s};d_{\s}^{+}\gg_{c}, \ll c_{k'\bs}^{+}d_{\bs}c_{k\s}; 
d_{\s}^{+}\gg_{c}$, and $\ll c_{k'\bs}^{+}c_{k\bs}d_{\s};d_{\s}^{+}\gg_{c}$ 
invovling two conduction electrons are neglected. After some algebraic 
manipulation the GF $\ll d_\s,d_\s^+\gg$ in the limit of $U\rightarrow\infty$ reads 
\begin{eqnarray} 
\ll&& d_{\s};d_{\s}^{+}\gg  =\left(1-n_{d\s} - V\sum_{k}\frac{\langle 
d_{\bs}^{+}c_{k\bs}\rangle}{\omega-\varepsilon_{k}}\right)(\omega-
\varepsilon_{d}+i\Delta \nonumber\\&& -V^2\sum_{k,k'}\frac{\langle 
c_{k'\bs}^{+}c_{k\bs}\rangle}{\omega-\varepsilon_{k}}+\sum_{k'}\frac{V^3}
{\omega-\varepsilon_{k'}} \sum_{k}\frac{\langle d_{\bs}^{+}c_{k\bs}\rangle}
{\omega-\varepsilon_{k}})^{-1}. \label{eq6} \end{eqnarray}
Eq.(\ref{eq6}) recovers exactly Lacroix's results. We shall discuss in
detail the solution of eq.(\ref{eq6}) together with the solutions of the next 
order approximation.

To go beyond the Lacroix's approximation it is necessary to consider the EOM of 
the connected GFs $\ll d_{\bs}^{+}c_{k'\bs}c_{k\s};d_{\s}^{+}\gg_{c}, \ll 
c_{k'\bs}^{+}d_{\bs}c_{k\s}; d_{\s}^{+}\gg_{c}$, and $\ll c_{k'\bs}^{+}c_{k\bs}
d_{\s}; d_{\s}^{+}\gg_{c}$, and to assume higher order correlation Green's 
functions to be zero. After a lengthy but straightforward calculation, $\ll 
d_{\s};d_{\s}^{+}\gg$ is  obtained finally as 
\begin{equation}
\ll d_{\s};d_{\s}^{+}\gg = \frac{1-n_{d\bs}-V\sum_k \frac{\langle d_{\bs}^{+}
c_{k\bs}\rangle } {\omega-\varepsilon_k}-\delta}{\omega-\varepsilon_{d}+i\Delta-
\sum_{k,k'}\frac{V^2\langle c_{k'\bs}^{+}c_{k\bs}\rangle }{\omega-\varepsilon_{k}}+
\gamma}, \label{eq7}
\end{equation}
with
\begin{mathletters}
\label{eq8}
\begin{eqnarray}
\delta=&&\frac{V^2}{n_{d\s}}\sum_{k}\frac{\langle d_{\bs}^+c_{k\bs}\rangle}{\omega-
\varepsilon_k}\sum_{k'}\frac{\langle d_\s^+c_{k'\s}\rangle}{\omega-
\varepsilon_{k'}}, \label{eq8a}\\ \gamma=&&-i\Delta\delta+\beta + 2\sum_{k,k'}
\frac{V^3\langle d_{\bs}^+c_{k\bs}\rangle}{(\omega-\varepsilon_k) 
(\omega-\varepsilon_{k'})} \nonumber\\&&+\frac{V^2}{n_{d\bs}}\sum_{k,k'} 
\frac{\left(\langle \hat n_{d\bs}c_{k'\bs}^{+}c_{k\bs}\rangle _{c} - 
\langle d_{\bs}^{+}c_{k\bs}\rangle  \langle c_{k'\bs}^{+}d_{\bs}\rangle \right)} 
{\omega-\varepsilon_{k}} \nonumber\\ &&  - \frac{V^3} {n_{d\s}}\sum_{k,k',k''}
\frac{\langle d_{\bs}^+c_{k\bs}\rangle \langle c_{k''\s}^+c_{k'\s}\rangle + 
\langle \hat n_{d\s}d_{\bs}^+c_{k'\bs}\rangle }{(\omega-\varepsilon_k)(\omega-
\varepsilon_{k'})} \nonumber\\ && + \frac{V^2}{n_{d\s}}\sum_{k,k'}\frac{\langle 
d_\s^+d_{\bs}^+c_{k'\bs}c_{k\s}\rangle - \langle d_{\bs}^+c_{k\bs}\rangle\langle 
d_\s^+c_{k'\s}\rangle}{\omega-\varepsilon_k}, \label{eq8b} \end{eqnarray}
where
\begin{equation}
\beta = \frac{V^2}{n_{d\s}}\sum_{k,k'}
\frac{\langle d_\s^+d_{\bs}^+c_{k'\bs} c_{k\s}\rangle_c - \langle 
d_\s^+c_{k'\bs}^+d_{\bs}c_{k\s}\rangle_c}{\omega- \varepsilon_{k}}. \label{eq8c} 
\end{equation} 
\end{mathletters}
In the above derivation, we have used the Hermitian conditions like $\langle 
d_{\bs}^+c_{k\bs}\rangle = \langle c_{k\bs}^+d_{\bs}\rangle$, etc. To simplify 
the solution and to obtain some analytical results, we notice
the following facts: ~({\bf i})~$\langle d_{\s}^{+}d_{\bs}^{+}c_{k'\bs}c_{k\s}
\rangle  \sim 0$ and $\langle \hat n_{d\s}d_{\bs}^{+}c_{k\s}\rangle \sim 0$ in 
the limit of $U\rightarrow \infty$, in which doubly occupied states are unfavorable.
 ~({\bf ii}) $\langle \hat n_{d\bs}c_{k'\bs}^+ c_{k\bs}\rangle $ is a new 
correlation function related to $\langle d_{\bs}^{+}\hat n_{d\bs}c_{k\bs} 
\rangle$ by the EOM and the spectral theorem. From observation ({\bf i}) it is 
evident $\langle d_{\bs}^{+}\hat n_{d\bs}c_{k\bs} \rangle\sim 0$. Thus, the 
spectral theorem implies that $\langle c_{k'\bs}^{+}\hat n_{d\bs}c_{k\bs}
\rangle\sim 0$. The cluster expansion yields simply $\langle \hat 
n_{d\bs}c_{k'\bs}^{+}c_{k\bs}\rangle _{c} - \langle d_{\bs}^{+}c_{k\bs}\rangle 
\langle c_{k'\bs}^{+}d_{\bs}\rangle \sim -n_{d\bs}\langle c_{k'\bs}^{+}c_{k\bs}
\rangle. $ ({\bf iii}) $\beta$ contains two higher order equal time correlations. 
One is the double hopping correlation $\langle d_\s^+d_{\bs}^+c_{k'\bs}c_{k\s}
\rangle_c$ which is approximately equal to $-\langle d_\s^+c_{k\s}\rangle\langle 
d_{\bs}^+c_{k'\bs}\rangle$ based on observation ${\bf(i)}$. The other is the 
spin-flip correlation $\langle d_\s^+c_{k'\bs}^+d_{\bs}c_{k\s}\rangle_c$, which 
should be calculated self-consistently. For simplicity, we just assume this 
correlation to be zero. Thus, we have approximately $$ \beta\sim- \frac{V^2}
{n_{d\s}} \sum_{k,k'}\frac{\langle d_{\bs}^+c_{k\bs}\rangle\langle d_\s^+c_{k'\s}
\rangle}{\omega-\varepsilon_k}. $$ {\bf iv)} By using the spectral theorem again, 
it is easy to show that
$\sum_{k'}V\langle d_\s^+c_{k'\s}\rangle=-i\Delta n_{d\s}$.

Based on  the above observations  we obtain finally \begin{eqnarray} \ll &&d_{\s};
d_{\s}^{+}\gg = \left(1-n_{d\s} - \sum_{k} \frac{V\langle d_{\bs}^{+}c_{k\bs}
\rangle} {\omega-\varepsilon_{k}}-\delta\right)\nonumber\\ &&\times(\omega-
\varepsilon_{d}+i\Delta(1-\delta) -2\sum_{k,k'}\frac{V^2\langle c_{k'\bs}^{+}
c_{k\bs}\rangle}{\omega-\varepsilon_{k}} \nonumber\\&&~~~~- \frac{1}{n_{d\s}}
\sum_{k,k',k''}\frac{V^3\langle d_{\bs}^+c_{k\bs}\rangle\langle c_{k''\s}^+c_{k'\s}
\rangle}{(\omega-\varepsilon_{k})(\omega-\varepsilon_{k'})})^{-1}, \label{eq9} 
\end{eqnarray}
where $\delta$ is given by  eq. (\ref{eq8a}).

The average functions $\langle c_{k'\bs}^{+}c_{k\bs}\rangle,~\langle c_{k''\bs}^{+}
c_{k'\bs}\rangle$ and $\langle d_{\bs}^{+}c_{k\bs}\rangle $ have to be calculated 
self-consistently. To distinguish our results for $\ll d_\s;d_\s^+\gg$ from 
Lacroix's we use $G_d^L(\omega+i\eta)$ for the Lacroix's solution and 
$G_d^N(\omega + i\eta)$ for our  solution. Following Lacroix, eq. (\ref{eq6}) 
reads\cite{la81} \begin{equation} G_d^L(\omega+i\eta)=\frac{1-n_d-A(\omega+i\eta)}
{\omega-\varepsilon_d +i\Delta+B(\omega+i\eta)-2i\Delta A(\omega+i\eta)}, 
\label{eq10} \end{equation} where \begin{mathletters} \label{eq11} \begin{eqnarray} 
A(\omega+i\eta)&=&-\frac\Delta\pi \int d\omega^{\prime}f(\omega^{\prime}) 
\frac{(\ll d_{\bs};d_{\bs}^{+}\gg)^{*}}{\omega^{\prime}-\omega-i\eta}, 
\label{eq11a} \\ B(\omega+i\eta)&=&\frac\Delta\pi \int d\omega^{\prime}
\frac{f(\omega^{\prime})}{\omega^{\prime} -\omega-i\eta}. \label{eq11b} 
\end{eqnarray} \end{mathletters} 
Similarly, the eq.(\ref{eq9}) becomes 
\begin{eqnarray} 
G_{d}^N(\omega+i\eta)=&&(1-n_{d}-A-\frac{A^2}{n_d})[\omega - 
\varepsilon_{d}+i\Delta(1-\frac{A^2}{n_d})\nonumber\\&& +2(B-i\Delta A)+
\frac {A}{n_d}(B-i\Delta A)]^{-1}. \label{eq12} 
\end{eqnarray} 
For the sake of simplicity, we have considered the nonmagnetic case, i.e. 
$n_{d\uparrow}=n_{d\downarrow}=n_{d}$ which is half of the total d-electron number 
$n_t$. By the spectral theorem: 
\begin{equation} n_t=\int f(\omega')\rho(T,\omega')d\omega^\prime, \label{eq13} 
\end{equation}
where $\rho(T,\omega) = -\frac2\pi\text{Im}G_d(\omega + i\eta)$ is the DOS with 
finite temperature and $f(\omega) = \frac{1}{\exp[(\omega - E_F)/T + 1]}$ is 
Fermi distribution function.

Eqs.(\ref{eq12}) and (\ref{eq10}) with (\ref{eq13}) constitute two sets of  
self-consistent equations respectively. They can be solved numerically. 
Before performing the numerical calculation we simply consider its analytic 
solutions in the high and low temperature limits. 

At high temperature $A(\omega+i\eta)$ is a small correction and thus an expansion 
with respect to $\Delta$ is available. After some algebra, one obtains the Kondo 
temperature 
\begin{equation} T_k = D\exp\frac{\pi(\varepsilon_d - E_F)}{\Delta}, \label{eq14} 
\end{equation} 
for the Lacroix's solution, and 
\begin{equation} T_k = D\exp\frac{\pi(\varepsilon_d - E_F)}{2\Delta}, \label{eq15} 
\end{equation} 
for our solution. Comparing with the exact expression obtained by Haldane
\cite{fh78} 
\begin{equation} T_k \approx (D\Delta)^{1/2}\exp\frac{\pi(\varepsilon_d - E_F)}
{2\Delta}, \label{eq16} \end{equation} our result is evidently an improvement. 
It is also helpful to check the low temperature result. At low temperature limit, 
as Lacroix noticed, $A(\omega+i\eta)$ must be large, especially as $\omega$ is 
near the Fermi level. We consider the zero temperature case. In this case, 
$A,B$ can be written as 
\begin{mathletters} \label{eq17} 
\begin{eqnarray} 
A(\omega+i\eta)&=&-\frac\Delta\pi G_d^*(\omega)\left(\frac{i\pi}{2} + 
\ln\frac{|\omega-E_{F}|}{D}\right),\label{eq17a} \\ B(\omega+i\eta)&=&\frac\Delta\pi 
\left(\frac{i\pi}{2}+\ln\frac{|\omega-E_{F}|}{D}\right). \label{eq17b} 
\end{eqnarray} 
\end{mathletters} 
Inserting $A$ and $B$ into the Eq.(\ref{eq12}), one obtain readily 
\begin{equation} G_d^N(E_F)=\frac\Delta\pi\frac{G_d^{N*}(E_F)}{2i\Delta^2/\pi 
G_d^{N*}(E_F) + \Delta/\pi}.        \label{eq18} 
\end{equation} The expression is exactly the same as that obtained by Lacroix.
\cite{la81}

In the following we solve numerically the DOS by the self-consistent equations 
(\ref{eq12}) and (\ref{eq13}) in the Kondo regime. The following parameters are 
considered. The total number of the d-electron is taken to be 0.9, which 
determines self-consistently the chemical potential.  The halfwidth $D$ is 
assumed to be 1, which defines the energy scale. The d-electron level 
$\varepsilon_d$ is taken to be 0 and the width of the virtual bound state 
$\Delta = 0.01D$. In Fig.\ \ref{fig1} the DOS for our solution (solid line) 
is shown at the very low temperature $T = 10^{-5}\Delta$. For comparison the 
DOS for the solution (10) with the above parameters (dash line), which is not 
available in Ref.\onlinecite{la82}, is also presented. In two cases the lorentzian 
resonance peaks are slightly shifted. In the Kondo regime better behaviors 
including larger $|\varepsilon_d - E_F|$ and a more pronounced Kondo peak 
are observed in our solution. A good test for the DOS is provided by the 
Friedel sum rule,\cite{frie52} which relates the DOS at the Fermi level to 
the occupation number $n_t$ at $T=0$ K, i.e.   $\rho(T=0,E_F) = 
\frac{1}{\pi\Delta}\sin^2(\frac{\pi n_t}{2}). $ \cite{costi94} This value is 
about 31 using the above parameter values. Our result of about 25 is in rough 
agreement with this exact value, while agreement is not as good for Lacroix's 
approximation. 

The temperature dependence of the DOS at the Fermi level is plotted in Fig.\ 
\ref{fig2} in the regime of the Kondo temperature, which is given by eq.\ 
(\ref{eq15}) in our calculation. It is evident that our result is comparable 
to the numerical renormalization group(NRG) result(dotted line). \cite{costi94} 
The result(dash line) from the Lacroix's approximation[eq.(10)] with the same 
parameters above is far below the NRG results. Therefore, the higher order 
correlation effects can not be neglected in describing the system properities  
quantatively.

In conclusion, by using the EOM for the connected GFs we obtained the solution 
of the SIAM beyond Lacroix's approximation. The missing factor of two in the 
expression of  Kondo temperature in Lacroix's approximation is regained, which 
is in agreement with the exact one except for a prefactor. The DOS and its 
temperature dependence at the Fermi level are obtained numerically. A  
quantitative improvement on Lacroix's approximation is found, which shows 
that the higher order correlation effects are important for describing the 
system properties quantatively.
\vskip0.06in

\noindent{\bf Acknowledgements} This work was supported in part by the
Deutscher Akademischer Austauschdienst (DAAD), and by the National Natural 
Science Foundation, the Doctoral Education Fund of Education Ministry, and the 
Nuclear Theory Research Center of HIRFL of China.

%\begin{thebibliography}{99}

\begin{figure}
\caption{ DOS for our solution(solid line) and for the Lacroix's approximation 
(dash line) at  $T = 10^{-5}\Delta$. The values of the parameters are  $n_d = 0.9$, 
$\varepsilon_d = 0.0$, $\Delta = 0.01D$, ~and ~$D = 1$. The d-electron level and 
the position for the Fermi level are indicated by the arrows.}
\label{fig1}
\end{figure}
\begin{figure}
\caption{Temperature dependence of $\rho(T,E_F)$. The parameters except the 
temperature are the same as those given in Fig.\ \ref{fig1}} \label{fig2} 
\end{figure}

\begin{references}
\bibitem{diss}  The temperature dependence of the DOS at the Fermi level was 
presented in Ref.\onlinecite{la82}. But the author used the parameter 
$\frac{E_0-E_F}{D} = 0.1$ ($E_0$ is the d-electron level), which does not 
describe the Kondo regime; according to Ref.\ \onlinecite{costi94}, his 
parameters describe the mixed valence or empty orbital regime.
\end{references}
\end{document}